\pdfoutput=1
\documentclass[11pt]{article}
\usepackage[affil-it]{authblk} 
\usepackage[authoryear]{natbib}
\usepackage{gensymb}
\usepackage{amsmath}
\usepackage{amssymb}
\usepackage{graphicx}
\usepackage{float}
\usepackage{booktabs}
\usepackage[a4paper, top=1in,bottom=1in,left=1in,right=1in,marginparwidth=0cm]{geometry}
\usepackage[title]{appendix}
\usepackage{pdflscape}
\usepackage{longtable}
\usepackage{setspace}
\usepackage{adjustbox}
\usepackage{array}
\usepackage{caption}
\usepackage{chngcntr}
\usepackage{mathtools}
\usepackage{lineno}
\usepackage{makecell}
\usepackage{hyperref}
\hypersetup{colorlinks=true,allcolors=blue}
\usepackage{physics}
\usepackage{bm}
\usepackage{textcomp, gensymb}

\newcolumntype{R}[2]{%
    >{\adjustbox{angle=#1,lap=\width-(#2)}\bgroup}%
    l%
    <{\egroup}%
}
\newcommand*\rot{\multicolumn{1}{R{45}{1em}}}
\newcommand{\citepos}[1]{\citeauthor{#1}'s (\citeyear{#1})}
\renewcommand{\eqref}[1]{Eq.\,\ref{#1}}

\newcommand{\E}[1]{\mathbb{E}\negmedspace \left[ #1 \right]}

\title{Weak signals, strong debates: Density dependence and population regulation through the lens of model uncertainty}

\author[1,*]{Evan C. Johnson}
\affil[1]{Naos Marine Laboratories, Smithsonian Tropical Research Institute, Ancón, Panama}
\affil[*]{Corresponding author: Evan C. Johnson, johnsone@si.edu}

\date{} 

\begin{document}

\maketitle 

\newpage

\tableofcontents 

\newpage

\section*{Abstract} 

Ecologists have long argued about the strength of density dependence and population regulation, respectively defined as the short-term and long-term rates of return to equilibrium. Here, I give three arguments for the intractability of population regulation. First, the ecological literature flip-flops on the strength of evidence for population regulation; by simple induction, population regulation should remain uncertain. Second, there is an analogous debate in economics about whether shocks to Gross Domestic Product have transient or permanent effects. This literature is extensive and sophisticated, yet there is no consensus, implying that more research will not resolve the issue in ecology. Third, using a variety of time series models and the Global Population Dynamics Database, I show that one's conclusions about population regulation are almost entirely dependent on model structure. This insurmountable model uncertainty explains why the strength of regulation is unresolved despite decades of research. However, it is possible to achieve the more modest goal of estimating density dependence. I introduce a novel measure of density dependence --- the effective autoregressive parameter --- which is conceptually intuitive and easy to calculate with simulations. The strength of density dependence varies significantly across populations, with an average that can be characterized as moderate: perturbations have an average half-life of 3 years. Rather than a universal balance of nature, stability varies widely across populations in ways that correlate with life history and taxonomy.

\newpage

\section{Introduction} \label{Introduction}

Ecologists have developed multiple definitions of stability based on different research contexts and ecological phenomena \citep{grimm1997babel, ives2007diversity}. Fredrick Clements defined stability as the long-term persistence of entire species assemblages (over ``thousands or even millions of years'' \citealp[p. 256]{clements1936nature}), a natural implication of his theory of climax communities. Charles Elton (\citeyear{elton1958ecology}) observed that cyclical dynamics were common in arctic communities, and defined stability as the absence of cycles or outbreaks. Robert Paine's (\citeyear{paine1966food}) rocky intertidal experiments, which showed how removing keystone predators led to prey species loss, led him to frame stability in terms of species coexistence. Yet another notion of stability is \textit{population regulation} --- the strength with which populations are pulled back toward their typical densities \citep{nicholson1933balance, lack1954natural, murdoch1994population}.

The concepts of density dependence and population regulation have been central to debates about ecological stability since the early 20th century. Nicholson argued that density dependence --- the relationship between population density and per capita growth rates --- was both common in nature and necessary for population regulation \citep{nicholson1933balance, nicholson1957self}. \citet{andrewartha1954distribution} showed that weather fluctuations could explain substantial variation in thrip population abundances, which seemed to diminish the importance of density dependence. This conflict came to a head at the 1957 Cold Spring Harbor Symposium, where eminent researchers debated density dependence with ``almost religious fervor'' \citep{macarthur1960review}. This debate and its resurgence in the 1990s (e.g., \citealp{wolda1995demise,  krebs1995two, turchin1995population}) is largely regarded as generating heat without light, probably due to slippery definitions of stability \citep{cooper2001must}.

Recent research has resolved the density dependence debate of the 1950s, but questions remain about the general strength of stability in natural ecosystems. Mathematical analysis has confirmed that density dependence is indeed necessary for population regulation \citep{chesson1982stabilizing, ellner1984asymptotic, hardin1988asymptotic}, and paradoxically, the explanatory power of the environment in Andrewartha's thrips is a sign of strong density dependence \citep{ziebarth2010weak}. However, these theoretical advances tell us nothing about the strength of density dependence and population regulation \textit{in general}. The overall level of stability has practical implications for predicting population abundances \citep[p. 29]{ziebarth2010weak} and optimizing restoration projects \citep{lampert2014optimal}. Such patterns may also inform the methodology of basic ecological research, e.g., weak regulation suggests a dominant role for stochasticity, thus favoring statistical-mechanics approaches like neutral theory and maximum entropy in predicting community patterns. Finally, these findings about stability speak directly to the popular notion of a \textit{balance of nature}; we return to this metaphor in the \hyperref[Discussion]{Discussion}.

\begin{figure}[H]
\centering
\includegraphics[scale = 1]{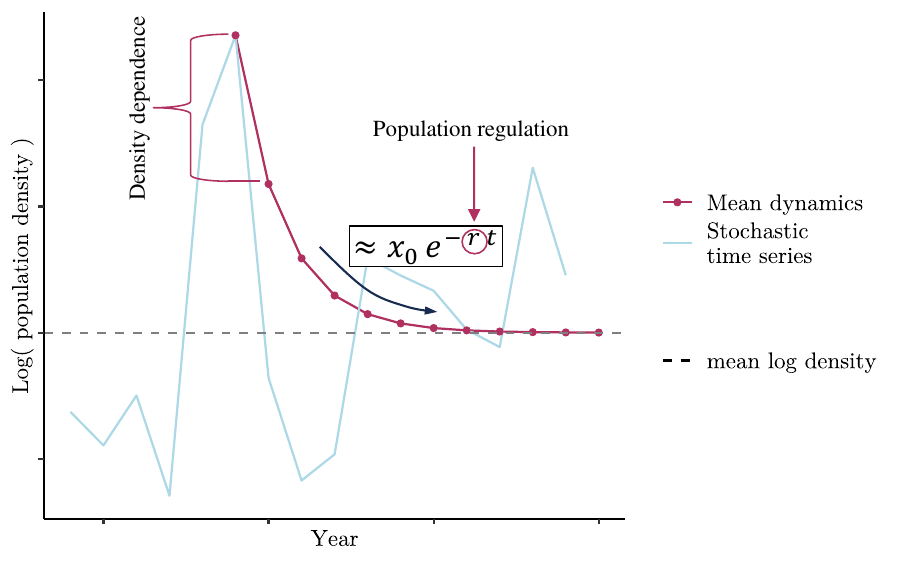}
\caption{A conceptual figure showing that density dependence and population regulation are respectively defined as short-term and long-term properties of ecological time series. Density dependence concerns the lag-1 relationship between density and growth rates. Population regulation concerns the asymptotic rate of return to equilibrium. Even though population densities are observed at discrete points in time, here I present the solution of an approximating continuous-time model, $dx/dt = -r x(t)$, to underscore the relationship between population regulation and the rate of return.}
\label{fig:concept}
\end{figure}

While often used interchangeably, density dependence and population regulation can be differentiated by their focus on short-term vs. long-term stability (Fig. \ref{fig:concept}; \citealp{hanski1990density, turchin1999population, berryman2002population}). Density dependence refers to the immediate effect of population size on per capita growth rates, while population regulation involves long-term stability, boundedness, or the return to a stationary distribution (i.e., a set of typical densities) following a disturbance. Density dependence is necessary for regulation but can also generate non-equilibrium dynamics such as cycles or chaos \citep{may1974biological}. Thus, there is no simple nor monotonic relationship between density dependence and population regulation. That being said, the distinction between these concepts becomes less important in certain contexts, such as autoregressive order-1 models where density dependence and regulation are equivalent (see the \hyperref[Methods]{Methods}).


There are conflicting lines of evidence for either weak or strong density dependence/population regulation (recall the two concepts are often used interchangeably). Evidence for weak density dependence includes 1) the relative rarity of population cycles and chaos, which are caused by strong density-dependent feedback loops \citep{hassell1976patterns, ellner1995chaos}; 2) statistical methods tend to overestimate density dependence in short time series \citep{kendall1954note, mcgilchrist1989bias}; and 3) the \textit{decoupled immigration} / \textit{asynchronous population trajectory} mechanism \citep{murdoch2003consumer, briggs2004stabilizing}, wherein strong density dependence in one life-stage is mitigated by dispersal, leading to weak intergenerational density dependence. On the other hand, evidence for strong density dependence includes 1) the fact that experiments usually detect density dependence (\citealp{thibaut2020hierarchical}, Appendix 1); 2) a lack of variation in population abundances leading to imprecision; 3) The \textit{statistical stabilization} mechanism, where apparent density independence results from averaging across sub-populations, due to the mismatch between the scales of sampling and dispersal \citep{de1991mobility, hastings1993complex, ray1996density}. While the aforementioned experimental studies are compelling, potential publication bias and disparate methdologies make it difficult to assess the strength of density dependence. A systematic analysis of multiple datasets is essential for addressing these uncertainties.

Here, I use a large database of ecological time series to estimate the general strength of density dependence and population regulation. I find that the strength of density dependence can be accurately estimated, whereas population regulation cannot. The discrepancy arises because population regulation reflects a long-term pattern, which is easily obscured by the accumulation of noise. Unless there is weak noise or hundreds of data points, model structure significantly influences the estimation process. Different models have different biases, and there is no good way to select or presuppose a particular subset of models.  Thus, one's largely subjective choice of model structure determines the strength of population regulation.

However, this paper's message is stronger than ``population regulation cannot be estimated with current methods'' ---  I contend that univariate time series data will never allow precise estimation of regulation \textit{in general}. While we cannot predict the future with certainty, historical patterns offer guidance. In ecology, the literature has failed to provide consensus about the strength of population regulation. A parallel exists in economics, where researchers remain divided on the persistence of shocks to the Gross Domestic Product (GDP).

The study of density dependence from time series data began in the 1970s, initially focusing on identifying cyclical and chaotic dynamics \citep{hassell1976patterns, berryman1999principles, ellner1995chaos}. More recent research has leveraged the Global Population Dynamics Database (GPDD), a collection of almost 5000 time series, to examine density dependence and population regulation directly. This cottage industry started with \citet{brook2006strength}, who used multiple models to infer that 92\% of populations showed strong evidence of density dependence. \citet{sibly2007stability} estimated return rates in the GPDD, but these results are likely unreliably due to technical difficulties with the \textit{theta-logistic model} \citep{doncaster2006comment, clark2010theta}. Next, \citet{ziebarth2010weak} found that many populations experienced weak population regulation. \citet{knape2012patterns} came to a similar conclusion, finding that observation error could give the illusion of strong density dependence. \citet{thibaut2020hierarchical} found that a hierarchical model structure increased evidence for density-dependent dynamics according to model selection procedures. Finally, \citet{gebreyohannes2024weak} analyzed an unprecedented 16,000 time series and found that density dependence did not significantly improve predictions.


This study addresses the shortcomings of previous time-series analyses with a large set of models, Bayesian methods, and a focus on estimation rather than hypothesis testing. Some of the aforementioned GPDD studies are framed as rebutting previous work, yet drawing direct comparisons is challenging because different studies use different methodologies. To resolve this, I use Bayesian methods to fit a large number of models (encapsulating many models used in previous studies) to a consistent subset of GPDD time series. Bayesian estimation provides a complete characterization of uncertainty through the posterior distribution, whereas maximum likelihood estimation only approximates uncertainty using the Hessian at a point estimate. This full accounting of of uncertainty is particularly valuable when posterior distributions are non-normal, and allows us to distinguish between-population differences from sampling error. Bayesian methods also allow for more reliable metrics of model fit \citep{vehtari2017practical}. Previous research often tested the null hypothesis of density independence --- a questionable choice given the wide consensus on the necessary existence of \textit{some} degree of density dependence \citep{murdoch1989analysis, godfray1992long, schreiber2012persistence}, rendering such tests redundant. Our analysis adopts \citepos{ziebarth2010weak} approach of estimating the magnitude of population regulation, presupposing that it exists in some capacity.


While only six previous studies have examined population regulation using the GPDD, economists have published extensively on an analogous research program. The question --- sometimes referred to as the unit root or trend stationary problem --- is whether shocks to Gross Domestic Product (GDP) are transient or permanent. Here, the shocks and transience are analogous to environmental noise and population regulation respectively. This sub-field of economics contains nearly 2000 studies (Fig. \ref{fig:econ}), which while not directly comparable to the aforementioned ecology studies (because not all of the econ studies utilize large datasets like the GPDD), does reflect a large research program with many datasets and a high degree of statistical sophistication; methods include non-parametric models, models for panel data, and fractional ARMA models which attempt to fit both short and long-term patterns. 

Despite extensive research, economists have not reached a consensus on the unit root problem \citep{diebold1996uncertain, libanio2005unit}. The original debate of whether shocks have transient versus permanent effects is unresolved, and it has been supplemented by debate on the tractability of the problem \citep{christiano1990unit, rudebusch1993uncertain, gordon1997stochastic}. There is further debate about the policy implications (e.g. \citealp{cerra2023hysteresis}), including how trend stationarity in the United States GDP would affect Barack Obama's chance at reelection (described in \citealp{cushman2016unit}). A prolific research program may risk becoming self-sustaining, where the justification for each new paper comes from the volume of prior studies  \citep{dennett2006higher} rather than the promise of progress.

In the following sections, I return to the primary argument: population regulation is difficult to estimate accurately because of insurmountable model uncertainty. To this end, I utilize the GPDD and fit Autoregressive (AR) models with various additional features, specifically moving average components, secular trends, non-normal process error, non-stationary process error, and observation error. To fully quantify model uncertainty, I integrate predictions across models using pseudo-bayesian model posterior probabilities.

\section{Methods} \label{Methods}

\subsection*{Data}

Time series data were obtained from the GPDD; \url{https://knb.ecoinformatics.org/view/doi:10.5063}). Time series were selected for analysis if the data were sampled annually, spanned at least 25 years, contained less than 20\% NA values, contained 5 or fewer consecutive NA values, and contained at least 5 unique positive values. The final collection of 1020 time series has a median of 38 observations (range: 25--157) and primarily represents mammals (44\%), insects (22\%), birds (18\%), and fish (14\%).

\subsection*{Models}

Population dynamics were modeled with autoregressive (AR) models with additional features. AR models can be derived as linear approximations of complex population dynamics \citep{gardiner1985handbook}, and are thus preferred when mechanistic, population-specific models are unavailable. The AR system evolves according to the equation
\begin{equation} \label{eq:AR}
x_t  =  \overline{x} + \sum_{i=1}^{P} \rho_i \left(x_{t-i} - \overline{x} \right)  + \epsilon_t,
\end{equation}
where $\overline{x}$ is the log-scale carrying capacity, the $\rho_j$ are autoregressive coefficients, and $\varepsilon_{t}$ represents process error, typically modeled as a normal random variate with standard deviation $\sigma_{\text{proc}}$. The state variable \(x\) is the logarithm of the population density. This transformation means that linear growth and additive noise in \(x\) become nonlinear growth and multiplicative noise on the scale of raw population density. To manage the issue of $\log(0)$ tending towards negative infinity, I adjusted the GPDD data by replacing zeros with the smallest non-zero value in each time series. NA values are handled by treating the missing population densities as latent parameters. I consider models with $P = 1, 2, 3, \text{or}\, 4$ autoregressive lags. Time-lags of a univariate time series contain information about unobserved variables \citep{deyle2011generalized}, so higher $P$ implies an ecological system with more state variables, e.g., competitors, natural enemies, age classes. 

To further expand on the set of possible models, I consider a range of additional features:

\begin{enumerate}
    \item \textbf{Moving Average Component}\\
    Moving average (MA) components can statistically adjust for observation error and temporally autocorrelated environmental noise \citep{staudenmayer2005measurement}, and have been used in previous studies of population regulation. A single MA term is added to the base model: \eqref{eq:AR} $ + \, \theta \varepsilon_{t-1}$, where $\theta$ is the first-order coefficient bounded by -1 and 1. While models with multiple MA lags were considered, they were not included due to model-fitting issues; specifically, multi-modal posteriors caused poor mixing.
    
    \item \textbf{Secular Trend}\\
    The log-scale carrying capacity becomes time-dependent and evolves according to $\overline{x}_t = \beta_0 + \beta_1t$. Trends are known to complicate the estimation of AR coefficients, particularly in short time series \citep{nelson1984pitfalls}. Random walks --- a particular type of no-regulation scenario --- naturally wander up and down, a pattern that can be mistakenly interpreted as a trend.

    \item \textbf{Non-normal Process Error}\\
    The normal distribution for process error $\epsilon_t$ is replaced with Student's t-distribution with an additional \textit{degrees of freedom} parameter, $\nu$. This parameter controls determines tail thickness, such that $\nu=1$ corresponds to a Cauchy distribution, and $\nu \rightarrow \infty$ corresponds to the normal distribution. Thicker tails allow for more extreme events, such as population crashes or mass immigration.
    
    \item \textbf{GARCH Process Error}\\
 Generalized AutoRegressive Conditional Heteroskedasticity (GARCH) models allow the process error variance to change over time. This may capture the intermittent volatility and cyclicity observed in some ecological time series (e.g., \citealp{henden2009nonstationary}). The typical parameterization is 
    \begin{equation}
    \sigma_{t, \text{proc}}^2 = \alpha_0 + \alpha_1(x_{t-1} - \overline{x})^2 + \alpha_2\sigma_{t-1, \text{proc}}^2. 
    \end{equation}
   The initial variance, $\sigma_{0, \text{proc}}^2$, is an additional parameter.
    \item \textbf{Observation Error}\\
   The observation error model adds Gaussian measurement error on the log-scale (a conventional scheme; \citealp{slade1977statistical, ives2010analysis, knape2012patterns}). The observations, denoted by $y_t$, are given by 
   $y_t = x_t + \eta_t$ with $\eta_t \sim \text{Normal}\left(0, \sigma_{\text{obs}} \right)$.

\end{enumerate}

I fit the cross-product of the 6 model structures (the baseline model plus the 5 additional features above) and the 4 AR lags, for a total of 26 models. All models were fit in a Bayesian manner, using Hamiltonian Monte Carlo as implemented by the \textit{Stan} program \citep{carpenter2017stan}. Using the methodology of \citet{jones1987randomly}, the AR parameters were given flat priors over the space of stationary solutions. More model-fitting details are provided in Appendix \ref{app:model_fitting}.

\subsection*{Measures of population regulation and density dependence}

Population regulation is tied to the \textit{stationary distribution}, a term of art for the long-run distribution of states in a bounded stochastic system. Here, ``Stationary'' refers to the fact that the moments of the distribution (e.g., mean, variance, skewness) do not change over time, and do not depend on initial conditions. The variance of the stationary distribution, denoted $\sigma_\infty^2$, characterizes the long-run variation in log population density. 

The strength of population regulation can be assessed through the dominant eigenvalue, $\lambda$, which determines how rapidly the system approaches its stationary distribution.  For example, the asymptotic dynamics of the mean is given by $\overline{x}_t - \overline{x}_\infty = \lambda \left(\overline{x}_{t-1} - \overline{x}_\infty\right)$.  When $\lambda$ takes complex values or negative real values, the system exhibits cyclical dynamics, with negative real values specifically indicating two-year cycles. The strength of population regulation is $1-\Vert \lambda \Vert$, which ranges from 0 (no regulation) to 1 (immediate regulation).


The \textit{effective autoregressive parameter}, $\rho^*$, measures the average shrinkage to carrying capacity in a single time step.
\begin{equation}
\rho^* = \E{\frac{x_t}{x_{t-1}}},
\end{equation}
where the expectation is taken over the stationary distribution of states. The strength of density dependence is then measured as $1- \abs{\rho^*}$. For AR models with 3 or fewer AR lags, $\rho^*$ can be computed analytically (see \textit{supplementary files}). For more complex models, $\rho^*$ can be estimated through simulation. Given a sufficiently long simulated time series of length $N$, we can use $\rho^* = (1/N) \sum{i=1}^{N} x_i/x_{i-1}$ from a long simulated time series. To improve the robustness of this estimate, using either a trimmed or winsorized mean is recommended, as this avoids extreme values that arise when the denominator ($x_{i-1}$) is close to zero. As demonstrated in supplementary figure \ref{fig:eff_rho_sim_vs_exact}, this simulation approach produces accurate results.

Population regulation can be computed using established techniques \citep{ives2003estimating}. The dominant eigenvalue is determined from the characteristic equation $0 = 1- \sum_{i=1}^P \rho_i z^i$, while the stationary distribution variance is obtained through matrix-vector algebra. These calculations simplify considerably for the special case of a first-order autoregressive model, i.e., 
$x_t = \rho_1 x_{t-1} + \epsilon_t$. In this case, $\lambda = \rho_1 = \rho^*$, and the stationary variance has the simple form $\sigma_\infty^2 = \sigma_{\text{proc}}^2 / \left(1-\rho_1^2 \right)$. This special case reveals two important principles: population regulation is determined by the autoregressive parameters rather than the error terms, and process noise interacts multiplicatively with regulation to determine the stationary variance. 

\subsection*{Dealing with model uncertainty}

Model uncertainty is explicitly propagated to our quantities of interest --- here, density dependence and population regulation. Given a single model, the Bayesian approach to quantifying uncertainty in some derived quantity is 1) draw samples of model parameters from the posterior distribution; 2), compute the quantity of interest for each parameter draw, thereby propagating parameter uncertainty; 3) display the resulting \textit{predictive posterior distribution} in some way, often a with a histogram. With multiple models, this procedure is repeated, but the number of posterior samples from each model is proportional to the \textit{posterior model probability}, which may be expressed using Bayes' theorem:
\begin{equation}
    \overbrace{P(\pmb{\bm{M}} | D)}^{\text{Model posterior probs.}} \propto \overbrace{L(D | \pmb{\bm{M}} )}^{\text{Marginal likelihoods}} \times \overbrace{P(\pmb{\bm{M}})}^{\text{Model priors}}. 
\end{equation}
Here, $\pmb{\bm{M}}$ is a vector of the 26 models, $D$ is the data, and $L(D | \pmb{\bm{M}} )$ denotes a vector of marginal likelihoods. The marginal likelihood is difficult to compute and has the undesirable property of being sensitive to prior specification \citep{gelman1995avoiding, dormann2018model}. Therefore, I take a conventional approach and replace $L(D | \pmb{\bm{M}} )$ with pseudo-bayesian model weights. The weight for model $i$ is $w_i = \exp(\text{ELPD}_i) / \sum_{j=1}^K \exp(\text{ELPD}_j)$, where $K$ is the number of models  \citep{geisser1979predictive, yao2018using}. ELPD stands for the expected log predictive density, which is a metric of overall model fit akin to AIC (Appendix \ref{app:model_fitting}). 

A typical choice of model prior is the discrete uniform distribution, i.e., $P(M_j) = 1/K$. However, this prior ignores the scenario where a single model best represents the data-generating process for most datasets. To explicitly model uncertainty in the model priors themselves, a hyperprior is placed on model probabilities (as in \citealp{ley2009effect, benavoli2013imprecise}), specifically a symmetric Dirichlet with the probability distribution function
\begin{equation}
P(\pmb{\bm{M}} = \pmb{x}) = \frac{\Gamma(K \alpha)}{\Gamma(\alpha)^K} \prod_{i=1}^K x_i^{\alpha}.
\end{equation}
The single concentration parameter is fixed at $\alpha=0.013$. This value was chosen so that 50\% of the prior mass corresponded to scenarios where $x_i > 0.9$ for some model. This half-sparse hyperprior distribution expresses the idea that a single model may be a systematically better representation of the true data-generating process. 


\section{Results} \label{Results}

To summarize the general strength of population regulation, I examined the proportion of populations whose posterior mean of $\Vert \lambda \Vert$ (or $\abs{\rho^*}$) exceeded specific threshold values. I defined strong and weak population regulation using the inequalities $\Vert \lambda \Vert < 0.5$ and $\Vert \lambda \Vert > 0.9$, which correspond to \textit{fluctuation half-lives} of approximately 1 and 6 years respectively. The fluctuation half-life represents how long it takes for a perturbation's influence to diminish by 50\%. We calculate this metric by solving the equation $\Vert \lambda \Vert^T = 0.5$ for $T$.

\begin{landscape}
\begin{figure}[H]
\centering
\begin{adjustbox}{margin=0pt 0pt 0pt -50pt} 
\includegraphics[scale = 0.85]{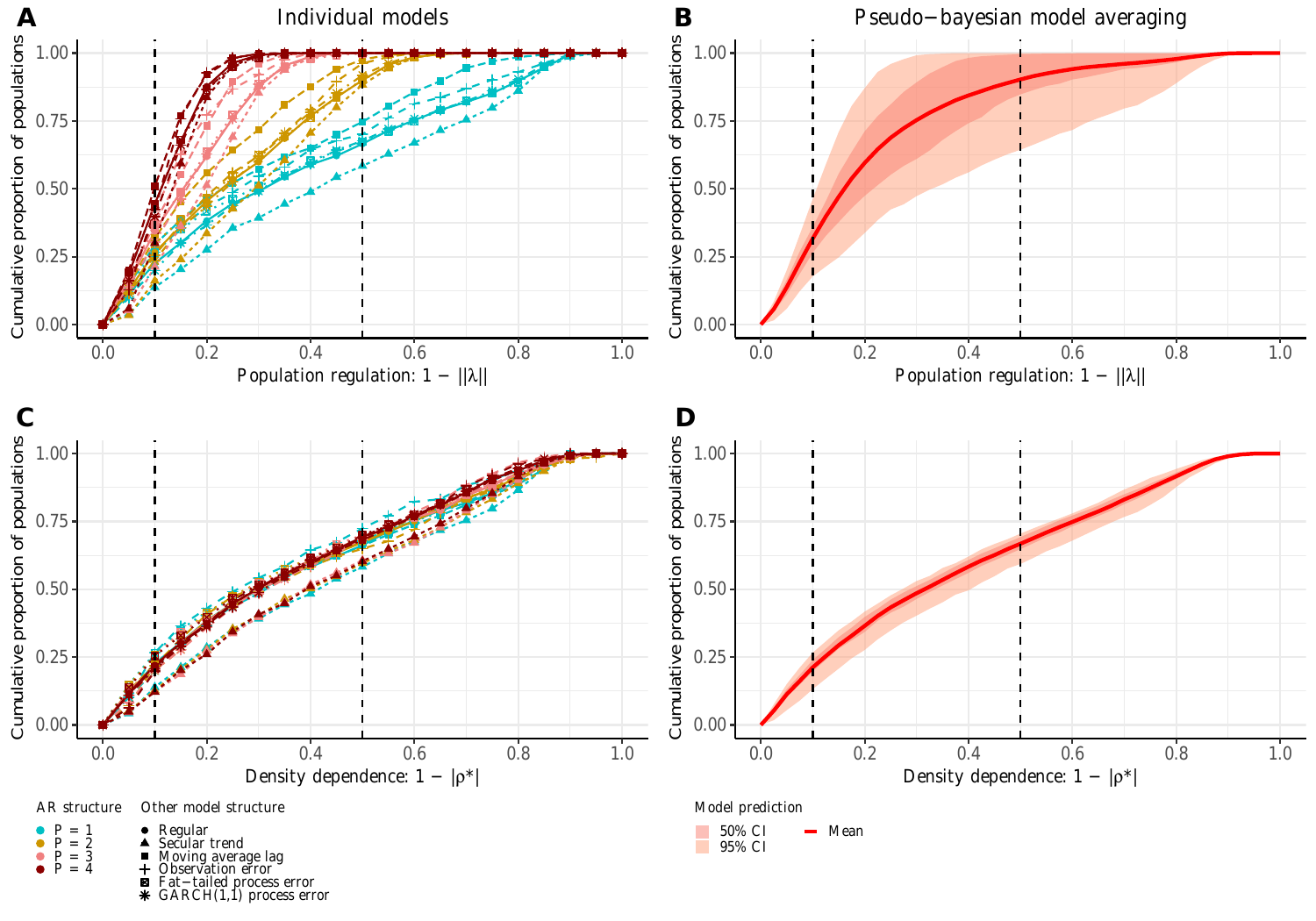}
\end{adjustbox}
\caption{There is large model uncertainty with respect to population regulation, but not density dependence. Each panel is the empirical cumulative distribution function across populations. The dashed vertical lines at 0.1 and 0.5 respectively show the thresholds, below and above which populations are considered to have weak or strong population regulation. Two different ways of dealing with model uncertainty are shown: 1) displaying models individually (panels A \& C), and integrating predictions across models with pseudo-bayesian weights (panels B \& D). Each population is represented by its posterior median of $\abs{\rho^*}$ or $\Vert \lambda \Vert$, i.e., within-population uncertainty is not represented.}
\label{fig:cdf_panels}
\end{figure}
\end{landscape}

Estimates of population regulation are highly dependent on model structure.  Roughly 12\% of populations are weakly regulated assuming an AR(1) model with a linear trend, whereas 50\% of populations are weakly regulated assuming an AR(4) model (Fig. \ref{fig:cdf_panels}, panel A). The situation is even worse when considering the frequency of strong regulation, which varies from 1\% to 45\% across model structures. Weighting by model-fit does not change this substantially: the Bayesian expression of model uncertainty shows that between 1\% and 30\% of populations could be strongly regulated (Fig. \ref{fig:cdf_panels}, panel B).

Estimates of density dependence are far more precise. Fig \ref{fig:cdf_panels} (Panels C \& D) shows that between 12 and 25\% of populations exhibit weak density dependence, and between 30 and 40\% of populations exhibit strong density dependence. The empirical cumulative distribution function of density dependence does not increase as sharply as that of population regulation, implying that the strength of density dependence varies substantially across populations.

There is no compelling justification for choosing a single model. For the majority of model comparisons, the disparity in model performance, as gauged by ELPD, is not statistically significant (Fig. \ref{fig:reject_hist}). Models with fat-tailed process error tend to perform better than other models (Fig. \ref{fig:ELPD_by_model}), but most of the uncertainty in population regulation and density dependence comes from models with differing numbers of AR lags. Longer time series favored models with more AR lags on average, but the difference was slight (Fig. \ref{fig:P_across_TS_length}).  

These findings are robust to a number of subjective modeling decisions. The substantial uncertainty in population regulation, which stems from model uncertainty, persists when analyzing only long time series (60 more years of data; Fig. \ref{fig:reject_hist_long}); or only high-quality time series (Fig. \ref{fig:cdf_panels_long_and_reliable}); when determining model parameters through Maximum Likelihood Estimation (Fig. \ref{fig:cdf_mle}); and when integrating uncertainty across models with Bayesian stacking weights (Fig. \ref{fig:cdf_panels_stacking}). The effective autoregressive parameter $\rho^*$ is not simply a surrogate for the lag-1 autoregressive parameter $\rho_1$. While these statistics are correlated (Fig. \ref{fig:rho1_eff_rho_scatter}), $\rho^*$ is conceptually distinct and is estimated with less model uncertainty (Fig. \ref{fig:rho1_cdf}).


\section{Discussion} \label{Discussion}

Population regulation is hard to estimate accurately because of \textit{model uncertainty} --- different models disagree about the strength of population regulation, and there is no basis upon which to select a single model. Density dependence can be estimated accurately, as all models are in rough agreement. The discrepancy between population regulation and density dependence can be explained by two facts. First, different model structures have built-in biases for population regulation, but not density dependence. Second, because regulation is a long-term property, its signal is eroded by the accumulation of noise across time steps.

\begin{figure}[H]
\centering
\includegraphics[scale = 1]{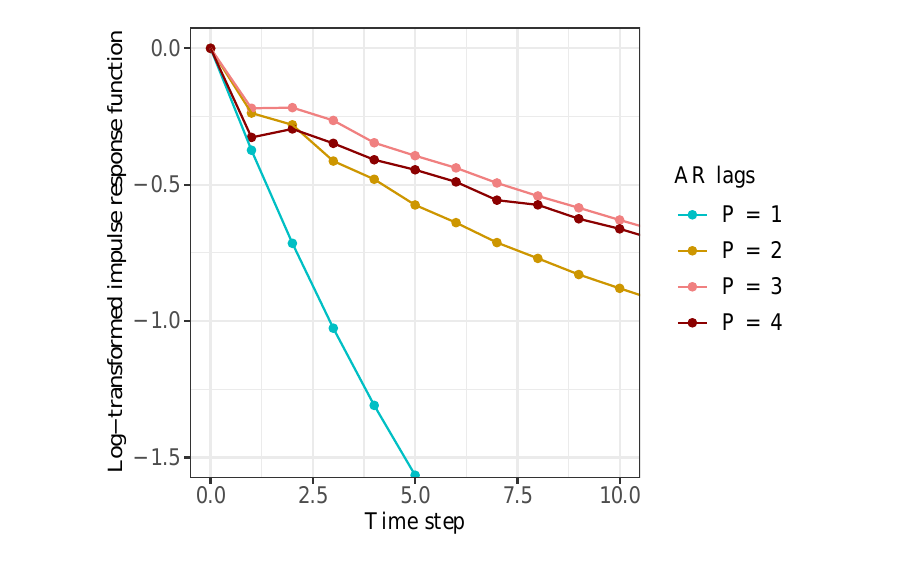}
\caption{All models make similar 1-year predictions, but different long-term predictions. The impulse response function (IRF) is the deterministic trajectory of log-abundance, starting one standard deviation above equilibrium. The slope of the log-IRF gives the asymptotic return rate. The deterministic trajectory is calculated by taking the mean across the posterior predictive distribution of trajectories. Only the baseline AR models are considered here.}
\label{fig:log_IRF}
\end{figure}

The divergence between short-term and long-term predictions can be visualized with an \textit{impulse response function}, which shows hypothetical noise-free trajectories with initial conditions 1 standard deviation above equilibrium. Figure \ref{fig:log_IRF}, shows the impulse response functions for models fit to a 63-year time series of the American black bear (GPDD ID: 106). All models produce a similar 1-year forecast, but more complex models predict a slow asymptotic return rate. This is likely a case of a more general phenomenon: ecological time series data, being typically short and noisy, make it challenging to estimate long-term properties, including future abundances \citep{beninca2008chaos,shaman2012forecasting}, Lyapunov exponents \citep{ellner1995chaos}, population trends \citep{white2019minimum}, and extinction probabilities \citep{fieberg2000meaningful}. To more systematically quantify how informative the data is relative to the model structure, I calculated the posterior contraction ($1 -\mathrm{Var}_{\text{prior}}/\mathrm{Var}_{\text{post}}$; \citealp{schad2021toward}) averaged across all populations (Table \ref{tab:post_contraction}). For density dependence, the posterior contraction is $\geq$ 0.67 across all models. For population regulation, the contraction is already 0.5 for $P=2$ and deteriorates to 0 for $P=4$. 


There is no good \textit{a priori} reason for selecting one model over another. Observation error is undoubtedly real, and has been estimated at around 10--20\% in nature \citep{ahrestani2013importance, francis2003quantifying, kamarainen2008zooplankton}; yet our statistical models may overestimate observation error by conflating it with process error, and the literature gives conflicting advice on whether or not to estimate observation error variance from time series data (\citealp{auger2021guide} vs. \citealp{ives2003estimating}). AR(1) models are parsimonious, but cannot model the substantial minority of series with cyclical dynamics \citep{louca2015detecting}. While ecosystems are typically viewed as complex systems (suggesting a large number of AR lags), many ecosystems can be well-approximated by low-dimensional systems \citep{ellner1995chaos, hsieh2008extending, munch2018nonlinear}. 

Both model selection and model averaging are problematic because they underreport model uncertainty. A workaround has been suggested by \citet[pg. 859]{ives2003estimating}: if population regulation is being estimated for many independent populations, and model selection leads to unbiased estimates of regulation, then the GPDD-wide mean of regulation will estimated with low bias (due to unbiased estimation) and low variance (due to averaging over many species). However, this approach led to substantial bias in a simulation study (Appendix \ref{sim_study}). I created simulated versions of the GPDD with autoregressive moving average (ARMA) models, and selected the best model for each population using sample size-corrected AICc. Comparing the true and estimated quantity of interest --- the proportion of populations exhibiting strong population regulation --- I found bias ranging from 0.02 to 0.2. The unpredictability of bias is relevant because it suggests that we can't perform a bias correction without already knowing the true model.

\begin{figure}[H]
\centering
\includegraphics[scale = 1]{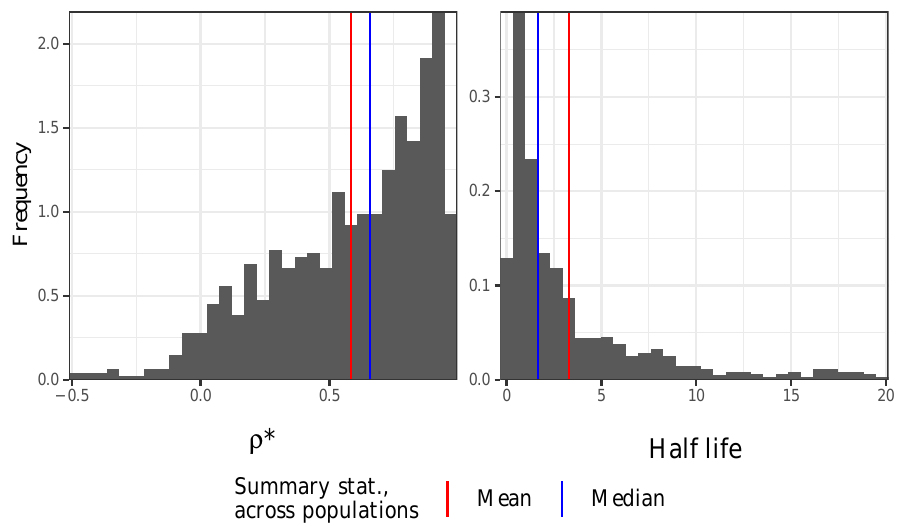}
\caption{Metrics of density dependence range widely across populations. Metrics include the effective autoregressive parameter, $\rho^*$, and the half life of fluctuations, calculated as $\log(1/2)/\log(\abs{\rho^*})$. Each population is represented by posterior medians. All models are weighted equally.}
\label{fig:hist_across_pops}
\end{figure}

How strong is density dependence? It is difficult to say, because the ``strong'' designation depends on one's prior expectations. That being said, an instructive metric is the \textit{fluctuation half life}, the number of years for a fluctuation to return half of the way to equilibrium on the log-scale, calculated by solving $0.5 =\abs{\rho^{*}}^{T}$ for $T$, assuming that repeated multiplication by $\rho$ adequately approximates the underlying deterministic dynamics. The median and mean of the of the population half-life are approximately 2 and 3  years (Fig. \ref{fig:hist_across_pops}). Subjectively, I would call this moderate density dependence. 

\begin{figure}[H]
\centering
\includegraphics[scale = 1]{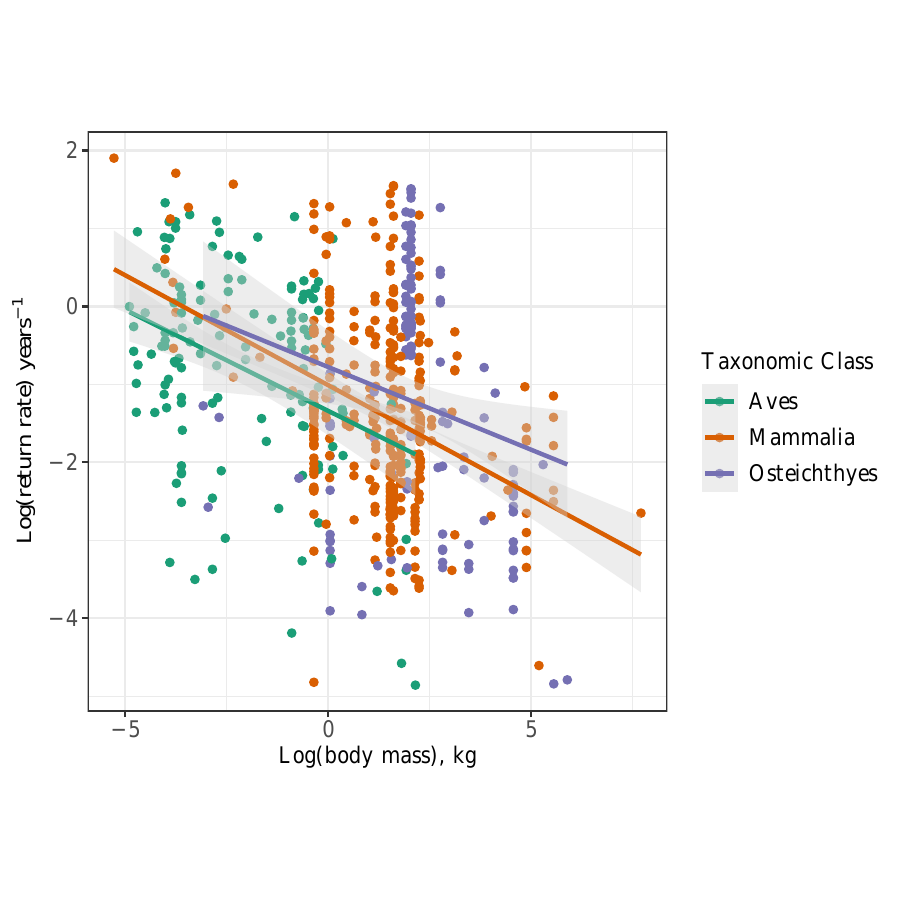}
\caption{Species with large body mass have weaker density dependence. The return rate is calculated as $\log(\abs{\rho^*})$. Body mass data comes from \textit{AMNIOTE} \citep{myhrvold2015amniote}, \textit{AnimalTraits} \citep{herberstein2022animaltraits}, and \textit{FishBase} \citep{froese2010fishbase} databases.}
\label{fig:body_mass}
\end{figure}

\begin{figure}[H]
\centering
\includegraphics[scale = 1]{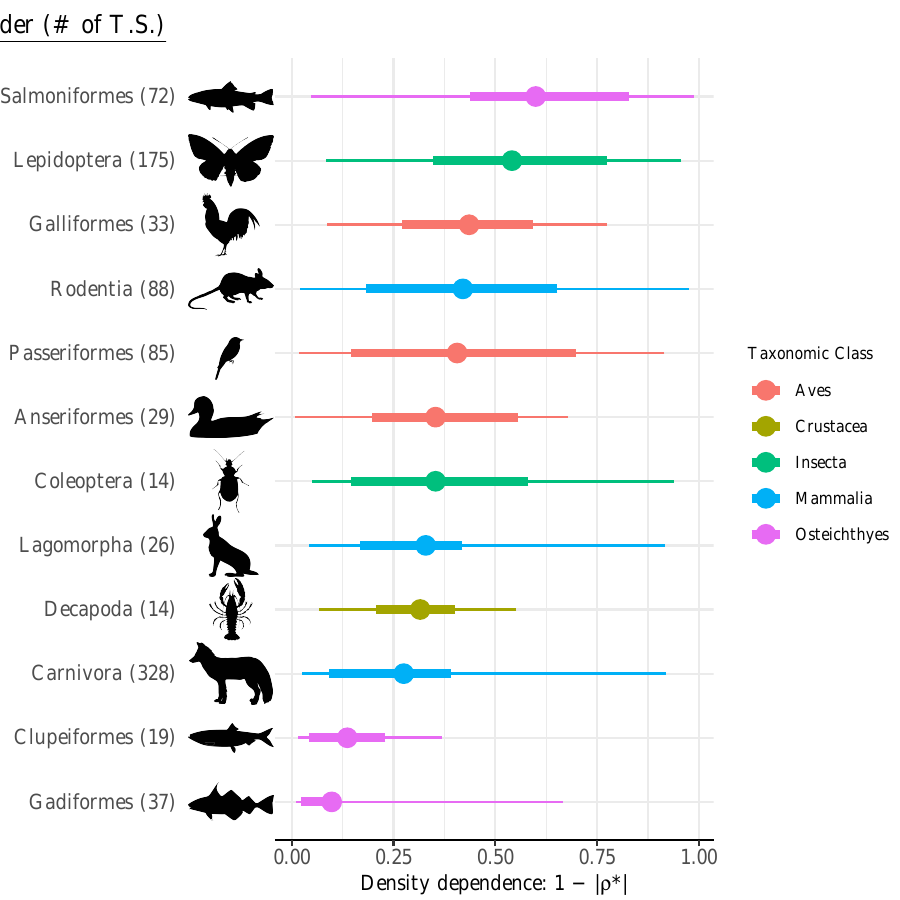}
\caption{Density dependence across taxonomic orders. Points, thick bars, and thin bars represent means, 50\% credible intervals, and 95\% credible intervals, across populations. Each population is represented by its posterior median, i.e., within-population uncertainty is not represented. Credits for silhouette images are provided in Table \ref{tab:silhouette}.}
\label{fig:eff_rho_by_order}
\end{figure}

Density dependence ranges widely across species. The interquartile range of $\rho^*$  is 0.37 to 0.85, which corresponds to a half-life range of 0.7 to 4.2 years. This variation cannot be explained purely by sampling error, since between population uncertainty is twice as large as parameter uncertainty (Table \ref{tab:var_partition}). Contrary to the findings of \citet{thibaut2020hierarchical}, I found that a hierarchical model structure did not reduce parameter uncertainty or dramatically alter the distribution of $\rho^*$ across species (Fig. \ref{fig:eff_rho_hierarchical_hist}; Appendix \ref{sec:hier}). The effective autoregressive parameter covaries predictably with body size (Fig. \ref{fig:body_mass}) and taxonomic order (Fig. \ref{fig:eff_rho_by_order}). Density dependence decreases with body size, likely due to slower maturation times, lower fecundities, and an increased capacity to buffer against bad weather. The strongest density dependence is observed among salmon species, which likely reflects the large and predictable swings in recorded abundance as salmon return to their natal streams. Within the \textit{Salmoniformes} order, the strength of density dependence correlates with the probability of cyclical dynamics (Fig. \ref{fig:salmon_cycles}). 

The intractability of population regulation is a minor loss for ecology, because the original reasons for studying population regulation have become less salient. The density dependence debates of the 20th century seem overwrought from a modern perspective, but may have reflected deeper anxieties about ecology's future. As \citet{cooper2001must} explains, ecologists feared that without density dependence, the pioneering work of Lotka \& Volterra would be irrelevant, and the integration of ecological with evolutionary biology --- a field which took density-dependent processes as the primary driver of natural selection --- would be impossible. If density dependence was not a general phenomenon, then the existence of any general phenomena was dubious, and thus, the status of ecology as a science was at stake; population ecology as a discipline was not firmly established until the 1950s, and theoretical ecologists were on the fringes until the 1960s \citep[p. 14--16]{kingsland1995modeling}.

But ecology has matured into a less anxious discipline, rendering questions of density dependence less critical. The fundamental question of whether density dependence is necessary for population regulation has been answered affirmatively \citep{hardin1988asymptotic, benaim2009persistence} and there is a general consensus that populations are regulated to some degree, at least at extremely low or high densities. More importantly, the discipline has expanded and adopted a pluralistic view of what qualifies as scientific. Traditional approaches like Lotka-Volterra models now coexist with newer methods such as neutral theory, maximum entropy theory, and bespoke population models. At the same time, the pursuit of general ecological knowledge has shifted from seeking universal laws to identifying generalities through meta-analyses, comparative case studies, and large-scale data analyses (like the one presented in this paper).

The concepts of density dependence and population regulation are perhaps most interesting in their connection to the balance of nature metaphor, which has shaped both scientific thought and popular understanding. The idea of balance has existed in antiquity \citep{egerton1973changing} and has persisted in popular culture, exemplified in Rachel Carson's \citeyear{carson1962silent} \textit{Silent Spring}, Ecology 101 concepts (e.g., trophic cascades, keystone species), and nature documentaries, which prefer to feature obligate mutualisms over dry discussions of functional redundancy. While ``balance'' is not strictly true in the sense that population abundances are constant, or that the removal of a single species would cause total food web collapse, there is nevertheless a sense that the web of causal relations between species will stabilize abundances \citep{mccann1998weak, hatton2024diversity}. This idea traces back to Darwin's strange metaphor where an ecological community is akin to a surface full of differentially shaped and rotated wedges \citep{darwin1859origin} --- The wedges (species) are in constant motion and therefore push against each other (interspecific interactions. When the room is saturated with wedges (niche packing), they eventually conform and form a stable interlocking mass. Tentative stability arises from competition and complexity.

Is there a balance of nature? We can answer this question only insofar as a single-species' density dependence is a reflection of a tentative balance arising from interspecific interactions (a dubious proposition; \citealp{cuddington2001balance}). While density dependence can be accurately estimated and is revealing of interesting biological patterns (e.g. Figs. \ref{fig:body_mass} \& \ref{fig:eff_rho_by_order}), there is not a dominant mode of density dependence, as measured with either $\rho^*$ or the fluctuation half-life. Given that populations experience a wide range of density dependence, the notion of nature's balance appears more complex than a simple yes or no --- as is often the case in ecology, the answer is ``it depends''.

\newpage


\section{Data availability statement}
Data and code are available on Zenodo (\url{TBD}).

\begin{appendices}
\counterwithin{figure}{section}
\counterwithin{table}{section}
\counterwithin{equation}{section}
\renewcommand{\thesection}{S\arabic{section}}

\section{Supplementary methods} \label{app:model_fitting}

Standard model-fitting diagnostics for Stan models were applied (\citealp[Ch. 6]{gelman2014bayesian}).  Models were removed from analysis if $\hat{R} > 1.1$ for any parameters (i.e., the chains had not mixed), if the effective sample size per iteration was smaller than 0.001 (i.e., inefficient sampling), if the energy Bayesian fraction of missing information (E-BFMI) was less than 0.2 (i.e., evidence for model misspecification), or if the proportion of divergent trajectories was greater than 1\% (i.e., evidence of biased estimation). This procedure led to the removal of 13\% of all models. Of these removed models, 94\% were models with observation error, and most had $P > 1$. Further investigation revealed that these models failed because of poor mixing and difficult posterior geometry (i.e., a high proportion of divergence trajectories), resulting in large $\hat{R}$ statistics.   

\begin{figure}[H]
\centering
\includegraphics[scale = 1]{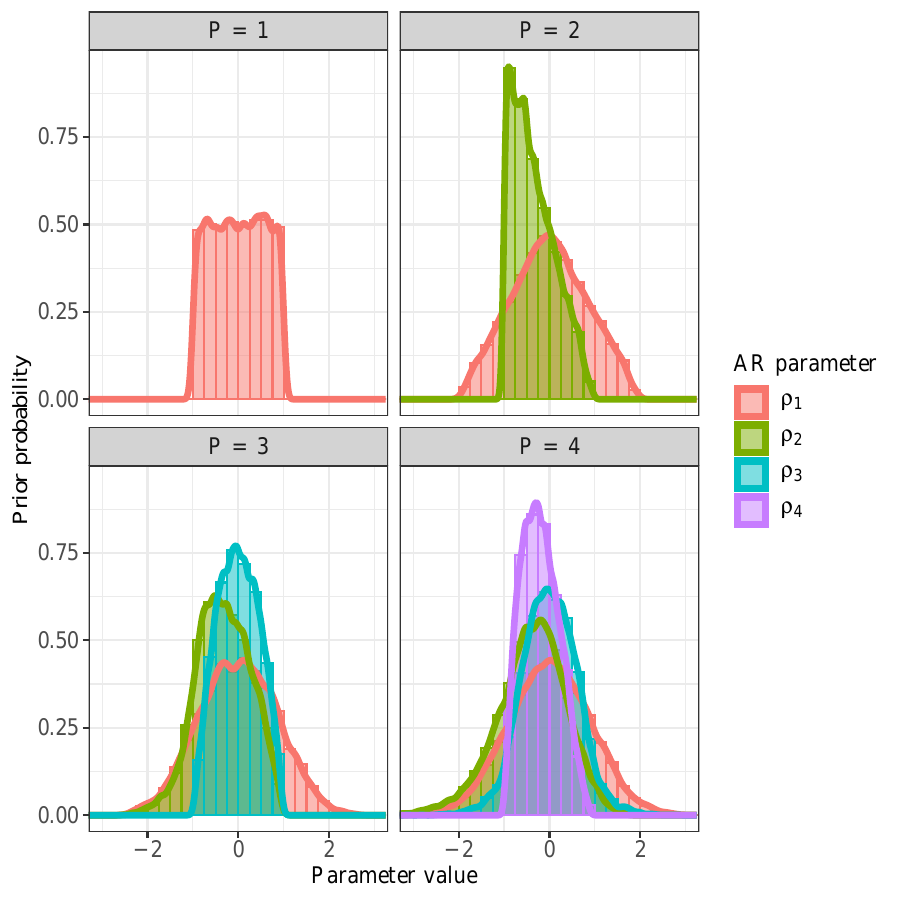}
\caption{Prior distributions of the autoregressive parameters.}
\label{fig:AR_prior_pdfs}
\end{figure}

The joint prior of the AR parameters must be specified with care to ensure that a stationary distribution exists. To this end, I applied a trick, first published by \citet{jones1987randomly}, in which partial autocorrelations are drawn from a beta distribution, and then jointly transformed into the usual AR parameters. The resulting joint prior distributions are flat over the space of stationary solutions (displayed in Fig. \ref{fig:AR_prior_pdfs}). 

Nearly all model parameters were given weakly informative prior distributions (\textit{sensu} \citealp{gelman2014bayesian}). The single exception was $\sigma_\text{obs}$, which was given a moderately informative prior in order to reduce the number of bad model-fitting diagnostics for the observation error models; the marginal prior was a half-normal distribution with scale parameter 0.2 $\times$ the standard deviation of log population densities. To monitor for an out-sized influence of prior choice on estimation, I computed the posterior contraction, $1 - \mathrm{Var}_{\text{prior}}/\mathrm{Var}_\text{post.}$, which measures the degree to which marginal posterior distributions are narrower than marginal prior distribution \citep{schad2021toward}. Posterior contraction was generally large for $\alpha_1$, $\alpha_2$, $\nu$, $\beta_0$, $\beta_1$, and $\sigma_\text{proc}$. Depending on model structure, the prior contraction could be small for the autoregressive parameters (see Table \ref{tab:post_contraction}), the moving average parameters, and the observation error standard deviation. When the observation-error model was fit successfully,  $\sigma_{\text{obs}}$ was often weakly non-identifiable with the standard deviation of process error $\sigma_{\text{proc}}$. This problem, the conflation of process error and observation error, is a common problem in ecological time series modeling \citep{knape2008estimability, thibaut2020hierarchical}.

Throughout the paper, I utilize a metric of overall model fit called the \textit{Expected Log Predictive Density} (ELPD). Conceptually, ELPD is the probability of a new dataset given old data, and thus implicitly contains a penalty for model complexity. The ELPD is asymptotically equivalent to the predominant information criteria for Bayesian models, the \textit{widely applicable information criterion} (WAIC). Additionally, ELPD provides several benefits over more popular information criteria --- unlike AIC and BIC, the ELPD integrates predictions over the posterior distribution and works with hierarchical model structures (like the observation-error models). ELPD can be estimated with Leave One Out Cross Validation (LOO-CV), along with a bias correction for the reduced sample size of the training data: $\text{ELPD}_\text{LOO-CV} = \left(\sum_{i=1}^N \log \text{Pr}(y_i | y_{-i})\right) + b$. Because LOO-CV is computationally expensive, I utilized an efficient approximation of LOO-CV using importance sampling, as implemented by the \textit{loo} package for \textit{R} \citep{vehtari2019loo}.

ELPD-based model weights likely understate model uncertainty because they penalize model complexity less severely than the true marginal likelihood. While the Bayesian Information Criterion (BIC) approximates the marginal likelihood \citep{ripley1992pattern} by applying a penalty proportional to log(sample size) $\times$ number of parameters, ELPD is asymptotically equivalent to AIC, which uses a smaller, fixed penalty. This lighter penalty in ELPD-based weights tends to favor more complex autoregressive models. The resulting pseudo-Bayesian weights are less balanced than those from the true marginal likelihood, effectively providing a lower bound on model uncertainty.

\section{Simulation study of model selection bias} \label{sim_study}

A simulation study was conducted to evaluate the bias and variance associated with a common model selection procedure for ARMA models. The study focused on a subset of ARMA models with autoregressive lags ranging from 1 to 3 and moving average lags of 0 or 1, resulting in six candidate models.

Ten different parameter sets were generated, each of which describes the population dynamics for the entire GPDD. To generate a single set, I drew model priors from a Dirichlet hyperprior distribution with concentration parameters $\alpha = 0.048$, chosen to achieve \textit{half-sparsity}:  50\% of prior mass is allocated to combinations where a single model receives 90\% probability. A larger $\alpha$ value was used here than in the main text's Bayesian analysis due to fewer models. Model posteriors were calculated using ELPD-based model weights, and model structures were selected using pseudo-Bayesian weights, computed as the product of model prior and ELPD-based weights. For each chosen structure, one posterior draw of model parameters was randomly selected. These steps were repeated 10 times to obtain 10 parameter sets, each representing a plausible data-generating process for the entire GPDD.

For each parameter set and each population, I simulated 30 time series that matched the length of the original GPDD time series. I then fitted all six ARMA models to each time series with restricted maximum likelihood (REML), using $R$ code from the supplementary materials of \citet{ives2010analysis}. I then selected the best model using the sample size corrected Akaike Information Criterion (AICc). From these fitted models, quantities of interest (QoI) --- the proportion of populations with $\Vert \lambda \Vert$ and $\rho^*$ above or below threshold values --- were calculated using the REML parameters. To assess performance, I calculated bias as the mean difference between QoI of the 30 simulated datasets and the true value. The precision was assessed with the standard deviation of QoIs across the simulated datasets.

The proportion of populations classified as having strong population regulation (i.e., $\Vert \lambda \Vert < 0.5$) had high precision (around 0.01), but an unpredictable bias, ranging from 0.0 to 0.16 across the 10 sets of parameters (Table \ref{tab:sim_study}). Similarly, the proportion of populations classified as having weak population regulation (i.e., $\Vert \lambda \Vert > 0.9$)  had a bias ranging from -0.11 to -0.19. The positive bias in Table \ref{tab:sim_study} represents a bias towards stronger density dependence/population regulation, which is a well-known result in time series analysis \citep{kendall1954note, sibly2007stability, ives2010analysis}.

The simulation study design achieved two objectives: the need for simulated data to establish ground truth for evaluation, and the importance of maintaining realistic parameter values by basing simulations on actual model fits to GPDD data. Several methodological simplifications were necessary to ensure computational feasibility: Time series were simulated with no missing values and the analysis was restricted to only 6 standard ARMA models. 

\begin{table}[H]
\centering
\begin{tabular}{lllllll}
\rot{Sim. parameter set} & \rot{True freq. of strong DD} & \rot{True freq. of strong PR} & \rot{Bias, freq. of strong DD} & \rot{Bias, freq. of strong PR} & \rot{SD, freq. of strong DD} & \rot{SD, freq. of strong PR}\\ 
  \hline
   1 & 0.320 & 0.174 & 0.073 & 0.049 & 0.010 & 0.009 \\ 
     2 & 0.321 & 0.034 & 0.094 & 0.157 & 0.009 & 0.011 \\ 
     3 & 0.320 & 0.244 & 0.070 & 0.024 & 0.009 & 0.009 \\ 
     4 & 0.321 & 0.136 & 0.080 & 0.109 & 0.010 & 0.013 \\ 
     5 & 0.327 & 0.278 & 0.060 & 0.011 & 0.008 & 0.009 \\ 
     6 & 0.337 & 0.232 & 0.080 & 0.081 & 0.009 & 0.012 \\ 
     7 & 0.307 & 0.136 & 0.088 & 0.075 & 0.008 & 0.010 \\ 
     8 & 0.327 & 0.218 & 0.088 & 0.081 & 0.009 & 0.010 \\ 
     9 & 0.309 & 0.170 & 0.086 & 0.079 & 0.008 & 0.011 \\ 
    10 & 0.332 & 0.269 & 0.093 & 0.064 & 0.010 & 0.014 \\ 
   \hline
\end{tabular}
\caption{Simulation study results. Rows index the ten parameter sets for the entire GPDD. ``DD'' stands for density dependence, and the frequency of strong DD is the proportion of populations with $\abs{\rho^*} < 0.5$. ``PR'' strand for population regulation, and the frequency of strong population regulation is the proportion of populations with $\Vert \lambda \Vert < 0.5$. Bias is the average difference between the quantity of interest of the simulated datasets and the true value, such that positive values represent a bias towards strong regulation. SD is the standard deviation of estimates across the 30 simulated datasets. } 
\label{tab:sim_study}
\end{table}

\section{Hierarchical AR(1) model details} \label{sec:hier}

A recent study found that a hierarchical model structure increased the evidence against the null hypothesis of density independence; at the same time, the hierarchical model structure greatly reduced the number of populations inferred to have strong density dependence, and reduced uncertainty in the GPDD-wide mean of density dependence (see \citealp{thibaut2020hierarchical}, Fig. 1). Motivated by these results, I constructed a hierarchical model to test the robustness of my results. 

To simplify the comparison between hierarchical and non-hierarchical models, I focused on the simple AR(1) model structure. To ensure stationary, I sampled the sole autoregressive parameter on the unconstrained scale $(-\infty, \infty)$, and then performed a transformation to constrain the parameter to the range $(-1, 1)$. 

Let $\phi_j$ be the unconstrained autoregressive parameter of population $j$. Note that the subscript no longer refers to the lag order, but rather the population index. The group-level mean and standard deviation, denoted $\mu_\phi$ and $\sigma_\phi$, are given weakly informative normal priors, 
\begin{equation}
\begin{aligned}
\mu_\phi &\sim \text{Normal}\left(0, 5\right)\\
\sigma_\phi &\sim \text{Normal}_{[0,\infty)}\left(0, 5\right)
\end{aligned}
\end{equation}
with the subscript ``$[0,\infty)$'' indicating that $\sigma_\phi$ has a half-normal prior. The population-level parameters are normally distributed: 
\begin{equation}
\phi_j \sim \text{Normal}\left(\mu_\phi, \sigma_\phi\right). \label{eq:pop_sample}
\end{equation}

Finally, to obtain the population-level autoregressive parameter, constrained to the range $(-1, 1)$, I apply the transformation
\begin{equation} \label{eq:constrain}
 \rho_j = -1+2 \times \frac{1}{1 + e^{-\phi_j}}.
\end{equation}
To construct an equivalent non-hierarchical version of this model, I simply replaced \eqref{eq:pop_sample} with a weakly-informative, zero-centered distribution, $\phi_j \sim \text{Normal}\left(0, 10\right)$, and then applied the constraining transformation defined by \eqref{eq:constrain}.

\begin{figure}[H]
\centering
\includegraphics[scale = 1]{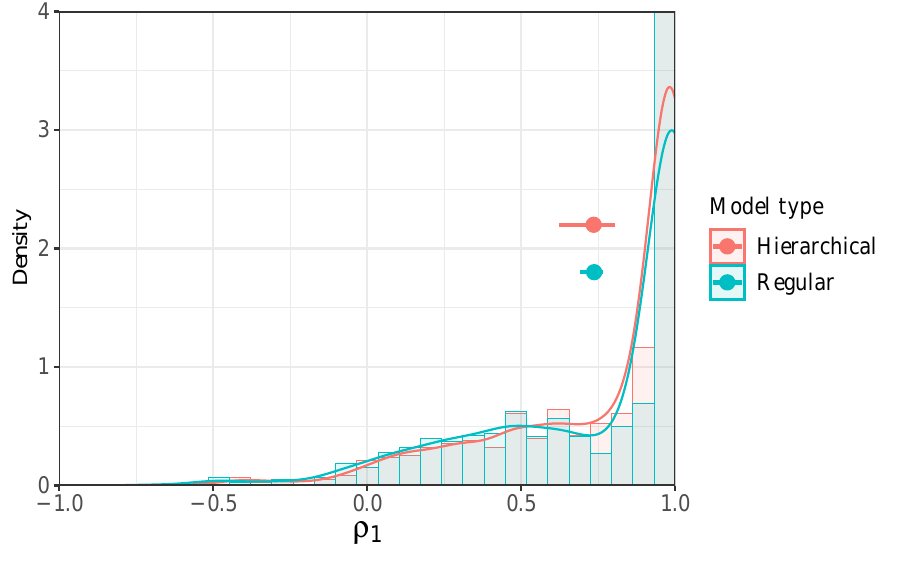}
\caption{A hierarchical model structure does not alter the estimated distribution of $\rho_1$ across populations. Each population is represented by its posterior median of $\rho_1$. The points and horizontal lines show the means and 95\% credible interval of the GPDD-wide mean of $\rho_1$.}
\label{fig:eff_rho_hierarchical_hist}
\end{figure}

\begin{figure}[H]
\centering
\makebox[\textwidth]{\includegraphics[scale = 1]{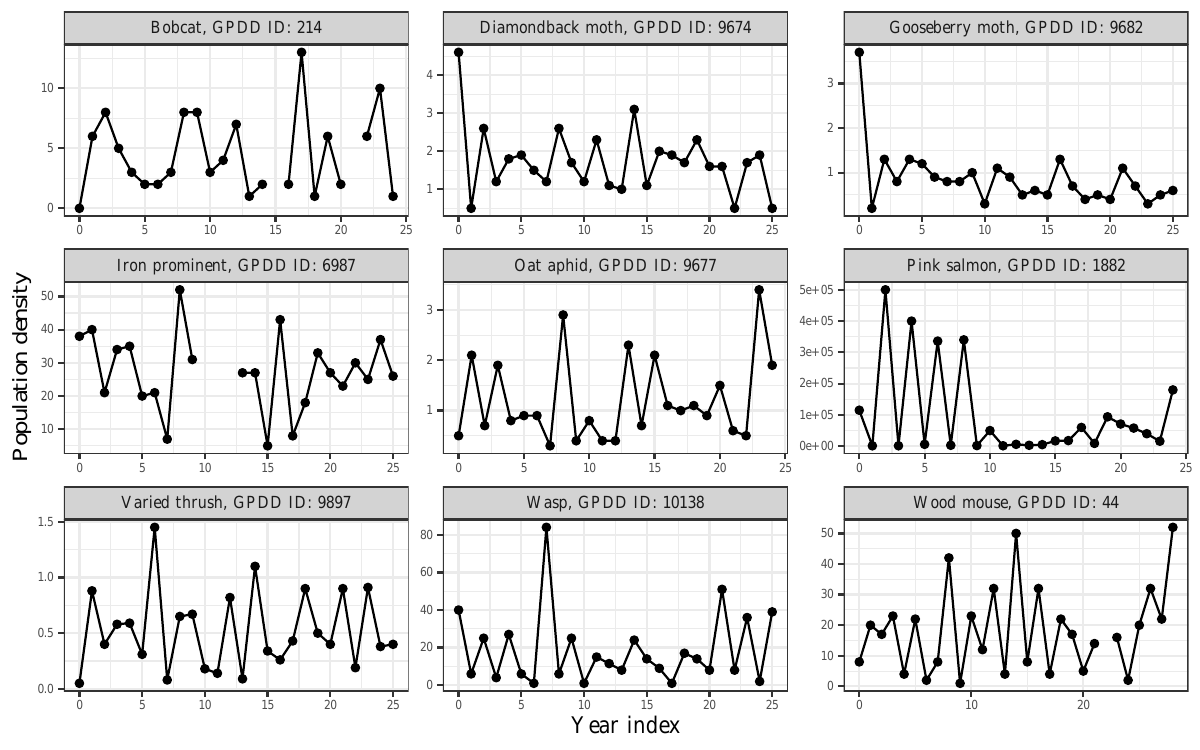}}
\caption{Some GPDD time series exhibit 2-year cycles. The populations displayed are the 9 populations (of distinct species) with the most negative posterior median values of $\rho_1$.}
\label{fig:negative_rho1_TS}
\end{figure}

The parameterized models show that the hierarchical structure does not alter the distribution of $\rho$ across species (Fig. \ref{fig:eff_rho_hierarchical_hist}), nor does it reduce population-level parameter uncertainty (Table \ref{tab:var_partition2}). It is difficult to isolate a primary reason for the difference between my results and the results of \citet{thibaut2020hierarchical}, given that there several differences between models: Thibault \& Connolly use a \textit{Gompertz} model structure, allow for non-stationary population dynamics, and fit the model with maximum likelihood. In favor of my results, I will cite 1) the robustness of Hamiltonian Monte Carlo (the algorithm used by the Bayesian model-fitting program \textit{Stan}), which is known to handle the curved geometry of likelihood surfaces posed by hierarchical models, and 2) the fact that my hierarchical model predicts a two-year population cycle (i.e., $\rho_j < 0$) for a small number of populations, whereas \citepos{thibaut2020hierarchical} hierarchical model does not (see their Figure 1). The two-cycle behavior can be clearly seen in some of the GPDD time series (Fig \ref{fig:negative_rho1_TS}).

\begin{table}[H]
\centering
\begin{tabular}{lll}
  & \multicolumn{2}{c}{Proportion of variance in...} \\ \cline{2-3}
Source of uncertainty & Hierarchical $\rho_1$ & Regular $\rho_1$ \\ 
  \hline
Within population & 0.32 & 0.23 \\ 
Between populations & 0.68 & 0.77 \\ 
\hline
\end{tabular}
\caption{A hierarchical model structure does not dramatically reduce within-population uncertainty, i.e., parameter uncertainty. Results were obtained by decomposing $\text{Var}(\rho_1)$ with the law of total variance.} 
\label{tab:var_partition2}
\end{table}

\section{Additional figures and tables}

\begin{figure}[H]
\centering
\includegraphics[scale = 1]{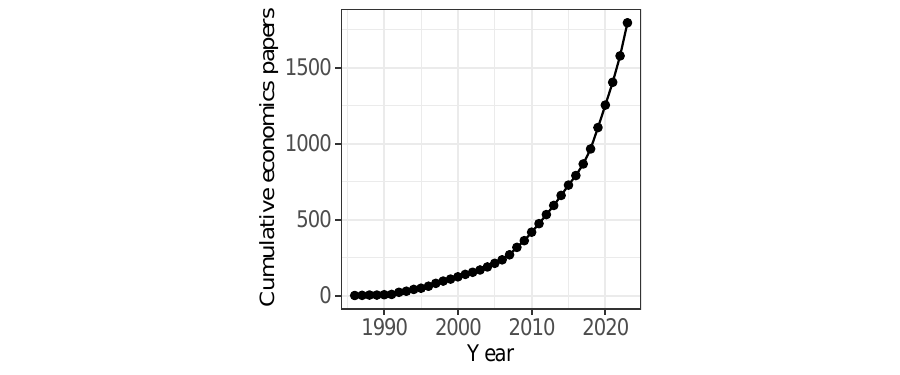}
\caption{The economics literature contains about 2000 papers on the topic of GDP shock transience (the econ-equivalent of population regulation). Data were collected from \textit{Web of Science} with the topic search, TS = (``GDP'' OR ``gross domestic product'' OR ``GNP'' OR ``gross national product'') AND (``unit root'' OR ``trend stationary'' OR ``difference stationary''). Items were further filtered to include only articles in business economics. A non-systematic review of search results confirmed that the vast majority of papers did indeed concern the problem of transience in GDP, also known as the \textit{unit root problem} or \textit{trend vs difference stationary problem}. The figure underestimates the true cumulative number of papers, since \textit{Web of Science} did not index abstracts or keywords before 1991. However, this underestimate is probably not severe, given that the seminal paper on this topic (\citealp{nelson1982trends}, now with over 8000 citations), was published less than a decade before the indexing began.}
\label{fig:econ}
\end{figure}

\begin{figure}[H]
\centering
\includegraphics[scale = 1]{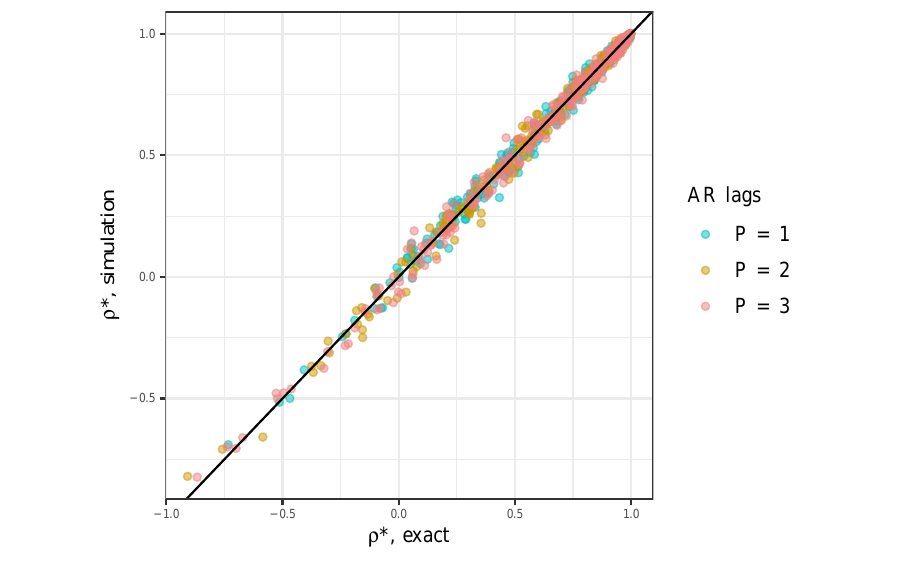}
\caption{The effective autoregressive parameter can be accurately estimated with simulations. The x-axis shows the exact $\rho^*$ for baseline AR models (see \textit{Mathematica} notebooks in the supplementary files). The y-axis shows $\rho^*$ via simulation, by taking the 5\% trimmed mean of $x_i/x_{i-1}$.}
\label{fig:eff_rho_sim_vs_exact}
\end{figure}

\begin{figure}[H]
\centering
\includegraphics[scale = 1]{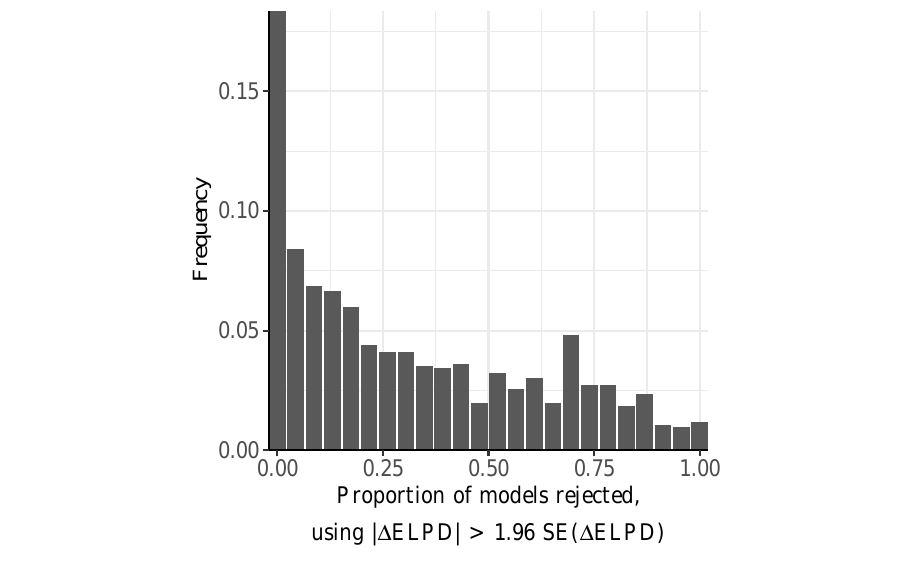}
\caption{Most models cannot be rejected, based on a significance test. The difference between models' ELPDs, denoted $\Delta \text{ELPD}$, is normally distributed, so $\abs{\Delta \text{ELPD}} > 1.96 \, SE(\Delta \text{ELPD})$ means that the null hypothesis $\Delta \text{ELPD} =0$ can be rejected at the 5\% confidence level.}
\label{fig:reject_hist}
\end{figure}

\begin{figure}[H]
\centering
\includegraphics[scale = 1]{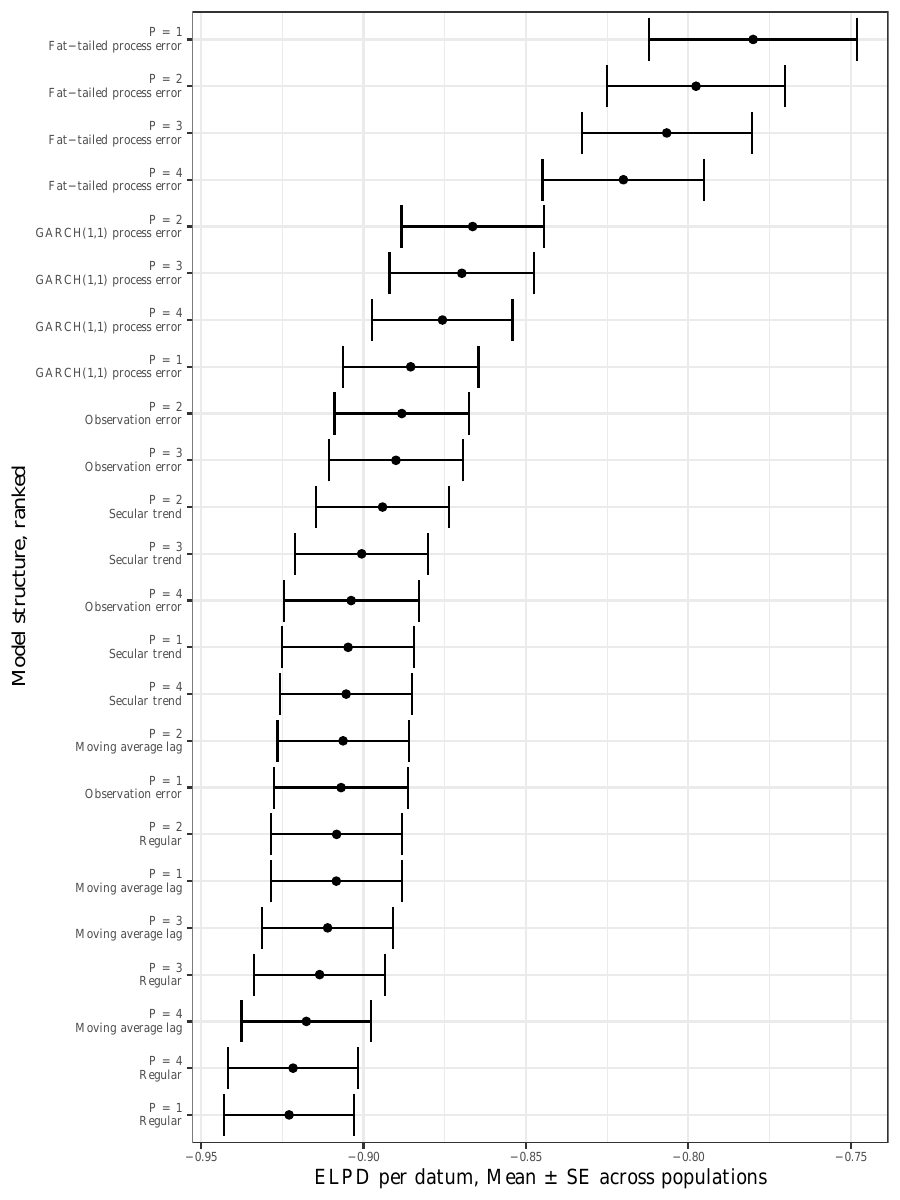}
\caption{Models with fat-tailed process error and GARCH error dynamics tend to fit the data better than other models.}
\label{fig:ELPD_by_model}
\end{figure}

\begin{figure}[H]
\centering
\includegraphics[scale = 1]{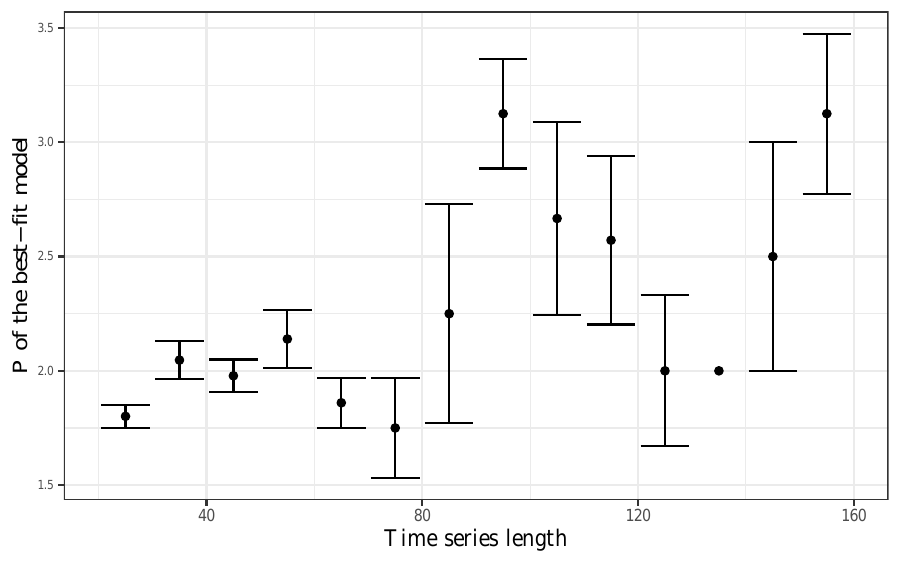}
\caption{The complexity of the best-fit model increases slightly as time series length increases. Model complexity is measured with the number of autoregressive lags; only baseline AR models are considered here. Model ``fit'' is measured with ELPD. The points and error bars represent the mean $\pm$ one standard error of P, across populations.}
\label{fig:P_across_TS_length}
\end{figure}

\begin{figure}[H]
\centering
\includegraphics[scale = 1]{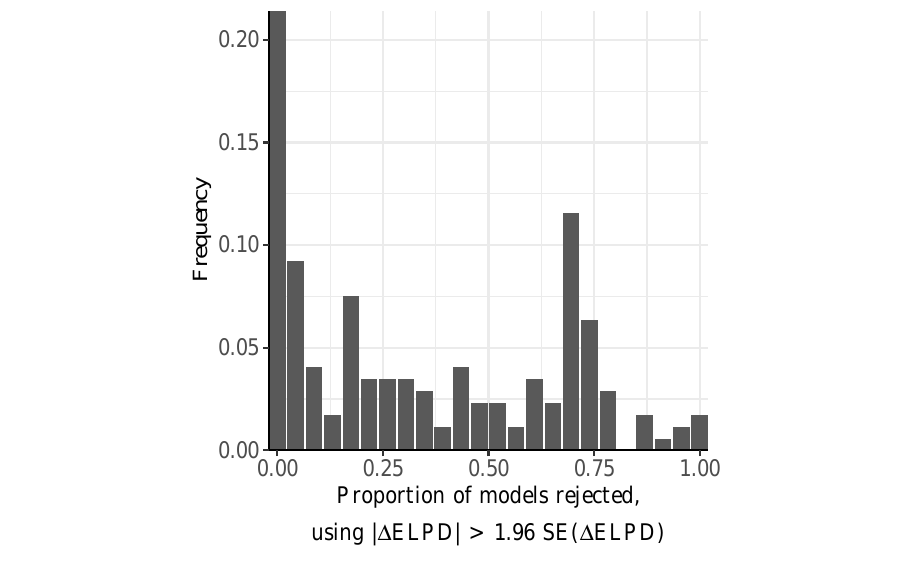}
\caption{Time series length does not dramatically alter the proportion of models that can be rejected at the 95\% confidence level. This figure is similar to Fig. \ref{fig:reject_hist}, except that only time series spanning 60 or more years (missing observations included) are considered. This filter drops the number of time series from 1020 to 173.}
\label{fig:reject_hist_long}
\end{figure}

\begin{landscape}
\begin{figure}[H]
\centering
\begin{adjustbox}{margin=0pt 0pt 0pt -50pt} 
\includegraphics[scale = 0.9]{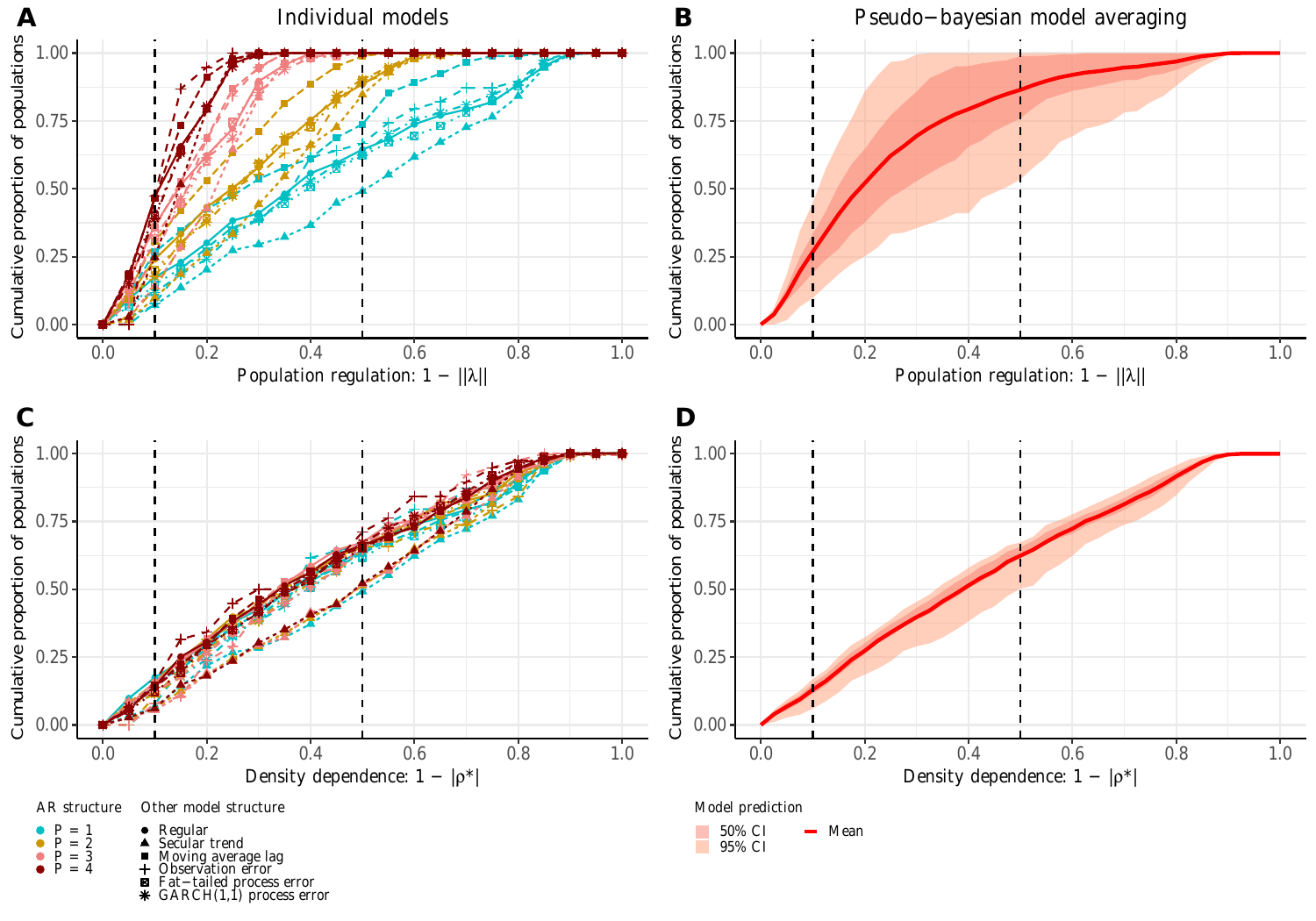}
\end{adjustbox}
\caption{The main results do not change when a particularly high-quality subset of the GPDD is used --- There is large model uncertainty with respect to population regulation, but not density dependence. This figure is similar to Fig. \ref{fig:cdf_panels} in the main text, except that GPDD series are only used if they have a reliability score $\geq$ 3 and ``Count'' based data (i.e., not harvest or area-based estimates). This filter drops the number of time series from 1020 to 183.}
\label{fig:cdf_panels_long_and_reliable}
\end{figure}
\end{landscape}
\newpage

\begin{figure}[H]
\centering
\includegraphics[scale = 1]{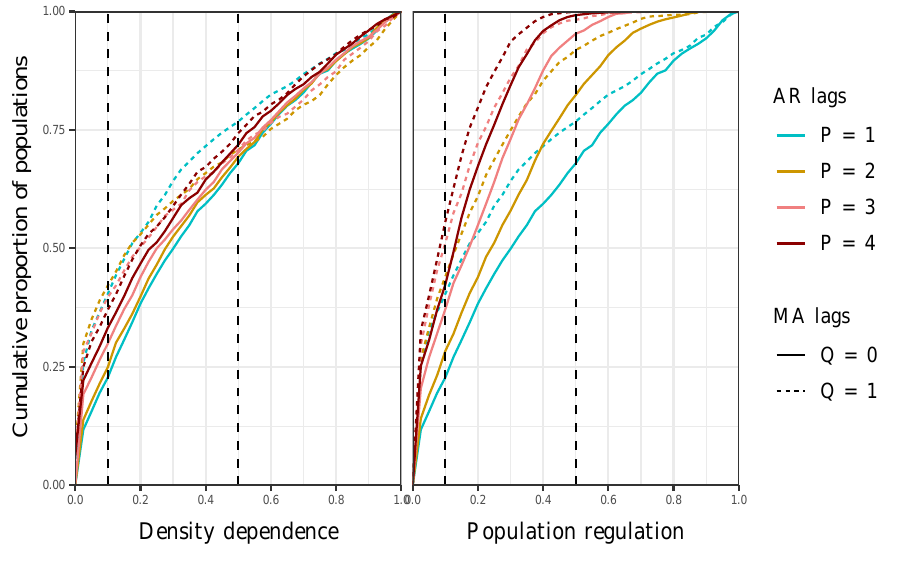}
\caption{Fitting the ARMA models with restricted maximum likelihood estimation (REML) does not alter the observed pattern of large between-model differences in population regulation.}
\label{fig:cdf_mle}
\end{figure}

\begin{landscape}
\begin{figure}[H]
\centering
\begin{adjustbox}{margin=0pt 0pt 0pt -50pt} 
\includegraphics[scale = 0.85]{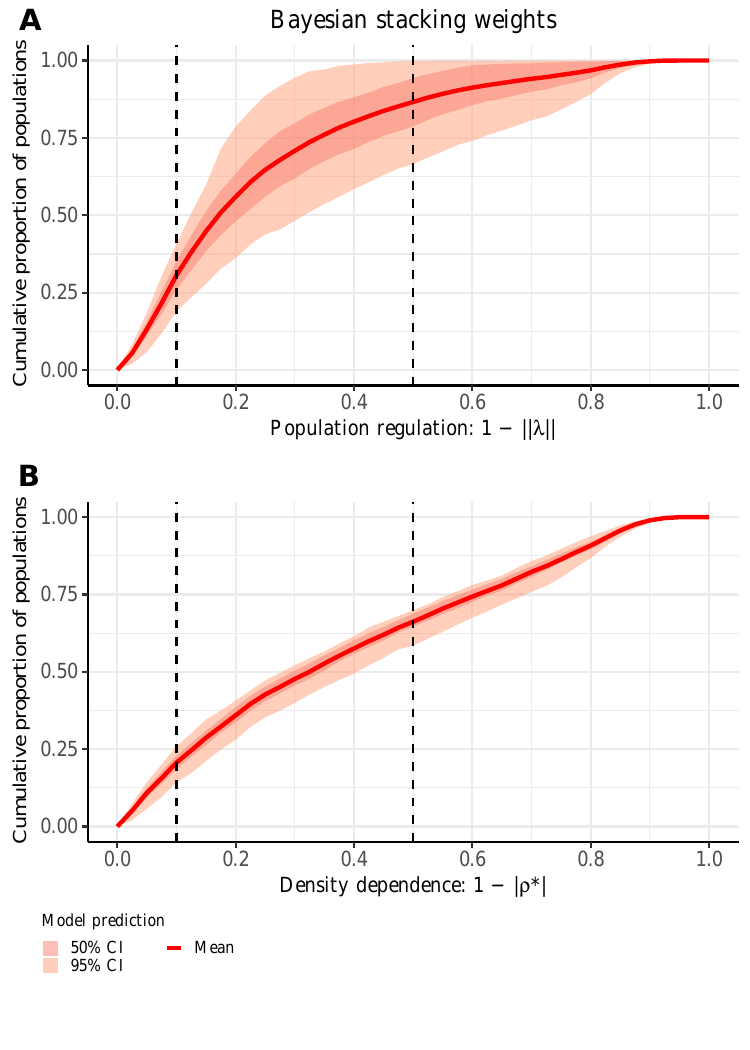}
\end{adjustbox}
\caption{Population regulation is more uncertain than density dependence when stacking weights are used to combine model predictions. Stacking weights are generated with the methodology of \citet{yao2018using} and the \textit{loo} package for \text{R} \citep{vehtari2019loo}. Panels A and B are analogous to Panels B and D in Figure \ref{fig:cdf_panels}, except stacking weights are used instead of pseudo-bayesian weights.}
\label{fig:cdf_panels_stacking}
\end{figure}
\end{landscape}

\begin{figure}[H]
\centering
\includegraphics[scale = 1]{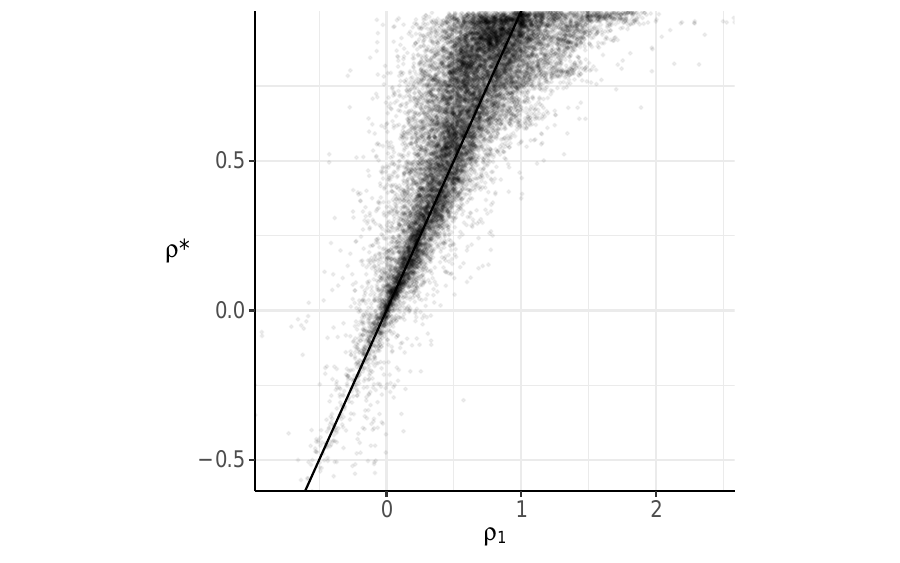}
\caption{The $\rho_1$ and $\rho^*$ statistics are correlated ($r = 0.77$), but are by no means identical, for the AR(2) and AR(3) model structures. The black line is the one-to-one line: $\rho^* = \rho_1$.}
\label{fig:rho1_eff_rho_scatter}
\end{figure}

\begin{figure}[H]
\centering
\includegraphics[scale = 1]{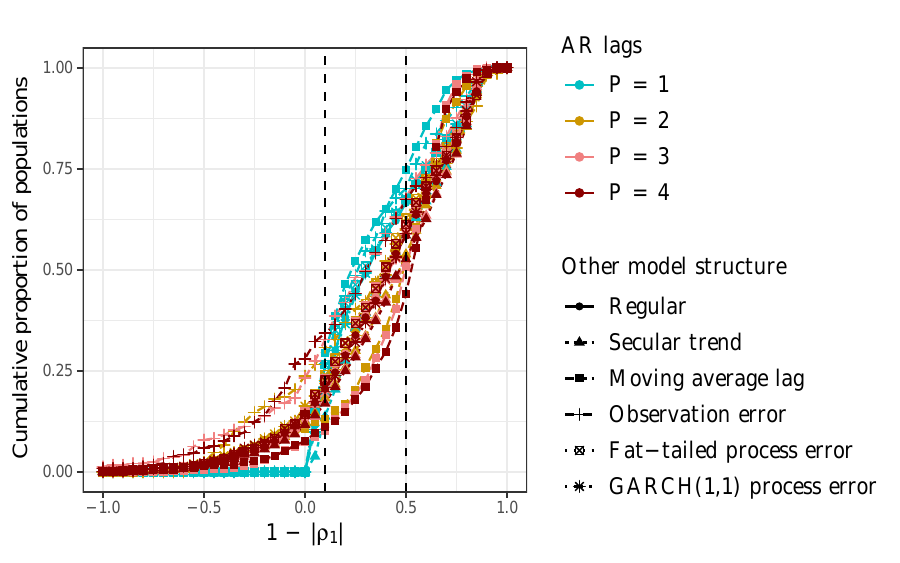}
\caption{The lag-1 autoregressive parameter, $\rho_1$, is estimated with more model uncertainty than the effective autoregressive parameter. Note that for AR models with $P>1$, a stationary distribution can exist if $\rho_1$ is greater than one; this explains negative values of $1-\abs{\rho_1}$.}
\label{fig:rho1_cdf}
\end{figure}

\begin{table}[H]
\centering
\begin{tabular}{lllll}
AR model structure & P = 1 & P = 2 & P = 3 & P = 4 \\ 
\hline
  $\rho^*$ prior mean & 0.51 & 0.50 & 0.49 & 0.51 \\ 
  $\rho^*$ post. mean & 0.63 & 0.64 & 0.64 & 0.64 \\ 
  $\rho^*$ post. shrinkage & 0.78 & 0.75 & 0.70 & 0.67 \\ 
  $\lambda$ prior mean & 0.51 & 0.75 & 0.86 & 0.91 \\ 
  $\lambda$ post. mean & 0.63 & 0.74 & 0.82 & 0.87 \\ 
  $\lambda$ post. shrinkage & 0.77 & 0.52 & 0.24 & -0.01 \\ 
   \hline
\end{tabular}
\caption{Time series data is weakly informative when estimating population regulation in autoregressive models with $P=3$ and $P=4$ time lags. Posterior contraction is measured by the relative precisions of marginal posterior and prior distributions: $1 - \mathrm{Var}_{\text{prior}}/\mathrm{Var}_\text{post.}$. Only baseline models are considered here, i.e., no MA terms, trends, observation error, or GARCH process error.} 
\label{tab:post_contraction}
\end{table}

\begin{table}[H]
\centering
\begin{tabular}{lll}
  & \multicolumn{2}{c}{Proportion of variance in...} \\ \cline{2-3}
Source of uncertainty & $\abs{\rho^*}$ & $\|\lambda\|$ \\ 
  \hline
Within population & 0.29 & 0.34 \\ 
  Between populations & 0.70 & 0.49 \\ 
  Between models & 0.01 & 0.17 \\ 
   \hline
\end{tabular}
\caption{Between-population differences account for most of the variation in density dependence (i.e., $1-\abs{\rho^*}$). Additionally, model uncertainty accounts for 17\% of the total variance in population regulation, but only 1\% of the total variance in density dependence. All 24 models are considered here.} 
\label{tab:var_partition}
\end{table}

\begin{figure}[H]
\centering
\includegraphics[scale = 1]{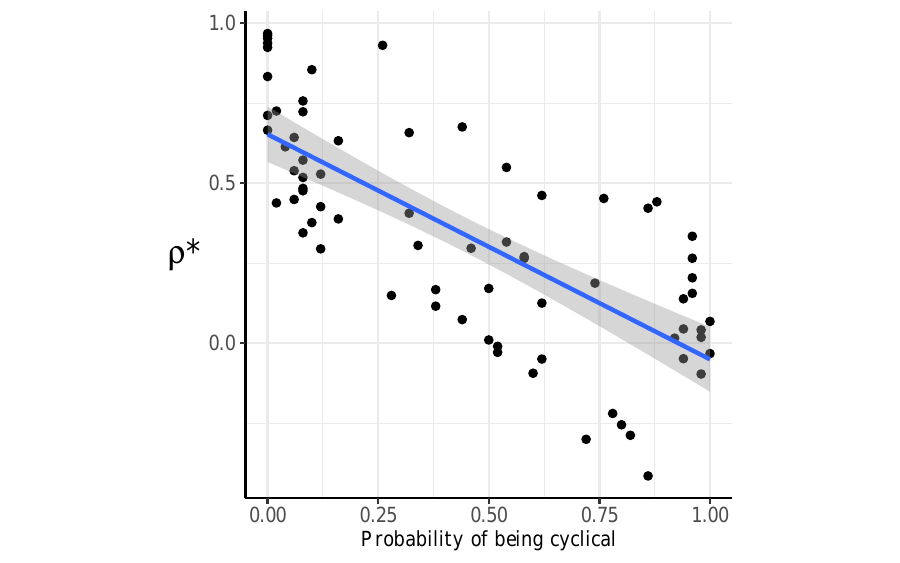}
\caption{Cyclical patterns are negatively correlated with density dependence in \textit{Salmoniformes} time series. Each point is a single 
population. For each posterior sample, I calculated the \textit{characteristic period} of the stochastic system (see \citealp{ives2010analysis}, pg. 861). Then, I calculated the population's probability of being cyclical as the proportion of posterior samples with a characteristic period that is finite and less than half the length of the time series. Each population's $\rho^*$ is a posterior median.}
\label{fig:salmon_cycles}
\end{figure}

\begin{table}[ht]
\centering
\begin{tabular}{p{0.2 \textwidth}p{0.4 \textwidth}p{0.4 \textwidth}}
  \toprule
  Taxonomic Order & Credit & License URL \\ 
  \midrule
Coleoptera & Prespa Research Group & \url{https://creativecommons.org/publicdomain/zero/1.0/} \\
Carnivora & Gabriela Palomo-Munoz & \url{https://creativecommons.org/licenses/by/4.0/} \\
Salmoniformes & Steven Traver & \url{https://creativecommons.org/publicdomain/zero/1.0/} \\ 
Passeriformes & Andy Wilson & \url{https://creativecommons.org/publicdomain/zero/1.0/} \\ 
Rodentia & Ferran Sayol & \url{https://creativecommons.org/publicdomain/zero/1.0/} \\ 
Gadiformes & Uncredited & \url{https://creativecommons.org/publicdomain/zero/1.0/} \\ 
Galliformes & Steven Traver & \url{https://creativecommons.org/publicdomain/zero/1.0/} \\ 
Lepidoptera & Didier Descouens (vectorized by T. Michael Keesey) & \url{https://creativecommons.org/licenses/by-sa/3.0/} \\ 
Lagomorpha & Ferran Sayol & \url{https://creativecommons.org/publicdomain/zero/1.0/} \\ 
Clupeiformes & xgirouxb & \url{https://creativecommons.org/publicdomain/mark/1.0/} \\  
Anseriformes & Andy Wilson & \url{https://creativecommons.org/publicdomain/zero/1.0/} \\ 
Decapoda & Pearson Scott Foresman (vectorized by T. Michael Keesey) & \url{https://creativecommons.org/publicdomain/mark/1.0/} \\ 
   \bottomrule
\end{tabular}
\caption{Silhouette images: credits and licenses} 
\label{tab:silhouette}
\end{table}

\end{appendices}

\newpage 

\bibliographystyle{apalike}
\bibliography{GPDD_refs.bib}


\end{document}